\documentclass[preprintnumbers,amsmath,amssymb,aps,prx,longbibliography,floatfix,lengthcheck,superscriptaddress]{revtex4-2}
\usepackage{units}
\usepackage{graphicx}
\usepackage{braket}
\usepackage{upgreek}
\usepackage{xcolor}
\usepackage{orcidlink}

\newcommand{\um}{$\upmu$m}

\begin{document}

\title{Coherent Control of the Fine-Structure Qubit in a Single Alkaline-Earth Atom}

\author{G. Unnikrishnan\orcidlink{0000-0002-7237-8462}}
\author{P. Ilzh\"{o}fer}
\author{A. Scholz\orcidlink{0009-0007-0522-6147}}
\author{C. H\"{o}lzl\orcidlink{0000-0002-2176-1031}}
\author{A. G\"{o}tzelmann\orcidlink{0000-0001-5527-5878}}
\author{R. K. Gupta\orcidlink{0000-0002-1380-0176}}
\author{J. Zhao}
\author{J. Krauter}
\affiliation{5. Physikalisches Institut and Center for Integrated Quantum Science and Technology, Universit\"{a}t Stuttgart, Pfaffenwaldring 57, 70569 Stuttgart, Germany}
\author{S. Weber\orcidlink{0000-0001-9763-9131}}
\author{N. Makki}
\author{H. P. B\"{u}chler}%
\affiliation{Institute for Theoretical Physics III and Center for Integrated Quantum Science and Technology, Universität Stuttgart, Pfaffenwaldring 57, 70569 Stuttgart, Germany}
\author{T. Pfau \orcidlink{0000-0003-3272-3468}}
\author{F. Meinert\orcidlink{0000-0002-9106-3001}}
\affiliation{5. Physikalisches Institut and Center for Integrated Quantum Science and Technology, Universit\"{a}t Stuttgart, Pfaffenwaldring 57, 70569 Stuttgart, Germany}
\date{\today}

\begin{abstract}
We report on the first realization of a novel neutral atom qubit encoded in the spin-orbit coupled metastable states ${^3\rm{P}_0}$ and ${^3\rm{P}_2}$ of a single $^{88}$Sr atom trapped in an optical tweezer. Raman coupling of the qubit states promises rapid single-qubit rotations on par with the fast Rydberg-mediated two-body gates. We demonstrate preparation, read-out, and coherent control of the qubit. In addition to driving Rabi oscillations bridging an energy gap of more than 17 THz using a pair of phase-locked clock lasers, we also carry out Ramsey spectroscopy to extract the transverse qubit coherence time $T_2$. When the tweezer is tuned into magic trapping conditions, which is achieved in our setup by tuning the tensor polarizability of the ${^3\rm{P}_2}$ state via an external control magnetic field, we measure $T_2 = \unit[1.2]{ms}$. A microscopic quantum mechanical model is used to simulate our experiments including dominant noise sources. We identify the main constraints limiting the observed coherence time and project improvements to our system in the immediate future. Our work opens the door for a so far unexplored qubit encoding concept for neutral atom based quantum computing.
\end{abstract}

\maketitle

Quantum computing with neutral atoms trapped in optical tweezer arrays has seen unprecedented progress in the past few years. This not only comprises recent advances in achieving high-fidelity single- and two-qubit gate fidelities \cite{Levine2018,Madjarov2020,Graham2022}, but also entails unique possibilities enabled by the matured field of optical tweezer technology. Besides versatile control over the qubit architecture geometry and dimensionality \cite{Endres2016,Barredo2016,Mello2019,Barredo2018}, recent achievements in scalable coherent atom shuttling have opened the door for engineering arrays with dynamical qubit connectivity, exploited for the demonstration of key aspects of logical qubit control and quantum error correction \cite{Bluvstein2022,Bluvstein2023}.

While a plethora of impressive results has been achieved using monovalent atoms, systems with more than one optically active electron, such as alkaline-earth metals or lanthanides \cite{Cooper2018,Norcia2018,Saskin2019}, provide new means of control to extend the quantum computing toolbox \cite{Lis2023,Huie2023,Ma2023}. This comprises recent demonstrations of new qubit encoding concepts in strontium and ytterbium atoms, relying either on the prominent ultra-narrow ${^1\rm{S}_0}$-${^3\rm{P}_0}$ optical clock transition \cite{Schine2022,Young2020} or on nuclear spin states in fermionic isotopes either in electronic ground or metastable excited states \cite{Jenkins2022,Ma2022,Barnes2022,Ma2023}.
    
\begin{figure}[t]
  \centering
    \includegraphics[width=\columnwidth]{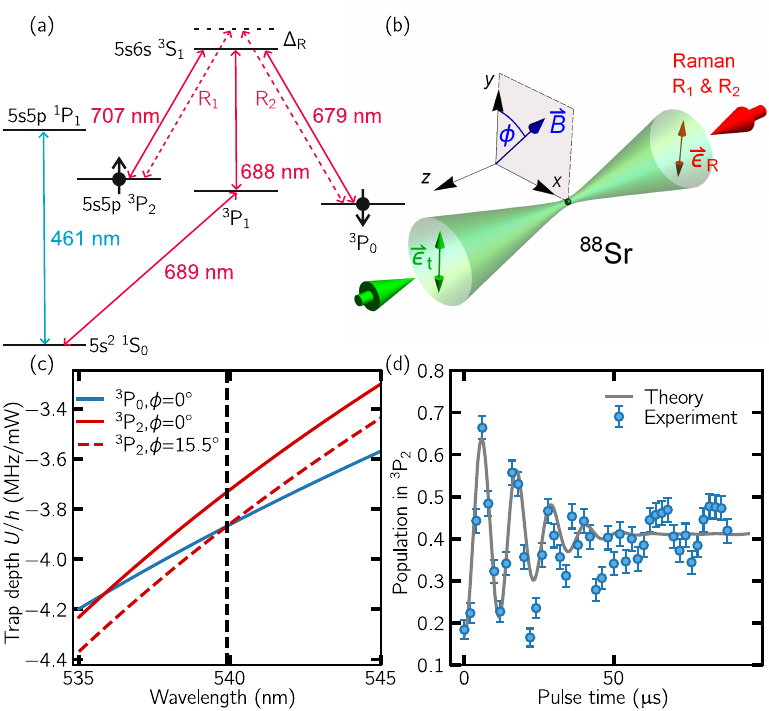}
    \caption{(a) Atomic level scheme of $^{88}$Sr depicting the fine-structure qubit in the states  ${^3\rm{P}_0}$ and  ${^3\rm{P}_2}$. Arrows indicate optical transitions relevant for this work. Single-qubit rotations exploit Raman coupling via $^3\rm{S}_1$ with single-photon detuning $\Delta_{\rm{R}}$ (dashed line). (b) Sketch of the experimental setup. Single $^{88}$Sr atoms are trapped in an optical tweezer (green). The Raman beams $\rm{R_1}$ and $\rm{R_2}$ (red) counter-propagate with the tweezer laser beam. The green (red) double arrow shows the linear polarization direction of the tweezer (Raman beams). An external magnetic field $\vec{B}$ in the $x$-$y$-plane allows us to tune the qubit magic wavelength via the azimuthal angle $\phi$. (c) Calculated trap depth for ${^3\rm{P}_0}$ and ${^3\rm{P}_2}$ as a function of tweezer wavelength for two values of $\phi$ \cite{SafronovaPC,Holzl2023}. The dashed line indicates the wavelength used in the experiment. (d) Rabi oscillation of the fine-structure qubit, showing the measured population in ${^3\rm{P}_2}$ as a function of pulse length of the Raman lasers. The tweezer power is $P=\unit[1.45]{mW}$ and $\phi = 0^\circ$. The gray line is a theoretical prediction including shot-to-shot fluctuations of the local Rabi frequency.
    }
    \label{Fig1}
  \end{figure}

Here, we report on the first realization of a third possibility for qubit encoding in divalent atomic systems. Specifically, we exploit the two long-lived fine-structure states ${^3\rm{P}_0}$ and ${^3\rm{P}_2}$ (magnetic quantum number $m_J=0$) of the metastable triplet $5\rm{s}5\rm{p}$-manifold in the bosonic $^{88}$Sr atom \cite{FSQubitPatent}. The two qubit states are gapped by $\unit[17.419]{THz}$ and are coupled using a two-photon Raman transition via the intermediate $5\rm{s}6\rm{s} \, ^3\rm{S}_1$ state (Fig.~\ref{Fig1}(a)). The fine-structure encoding comes with a number of notable benefits. First, the two-photon coupling promises single-qubit rotations on the \unit[100]{ns} timescale, orders of magnitude faster than what has been demonstrated on the comparatively slow ${^1\rm{S}_0}$-${^3\rm{P}_0}$ optical clock qubit, bringing single-qubit gate times on par with the fast Rydberg-mediated two-qubit gates. Compared to sub-microsecond gates realized in alkali atoms via Raman coupling utilizing vector light shifts \cite{Levine2022}, our qubit yields favorable heating properties for very large single-photon detunings, similar to recent explorations of lanthanides for spin-orbit coupled Fermi gases \cite{Burdick2016}. Second, the small momentum transfer associated with the Raman drive substantially reduces recoil heating and provides clear advantages for coherent qubit transport. Third, the qubit features magic wavelength trapping near \unit[540]{nm}, for which also Rydberg states are trapped via the polarizability of the Sr$^+$ ionic core. Furthermore, exploiting the tunability of the tensor polarizability of the ${^3\rm{P}_2}$ state via an external magnetic field allows for finding a triple-magic trapping scenario, for which the two qubit states and the Rydberg level experience equal ac-Stark shifts, providing ideal situations for immediate next steps toward high-fidelity two-qubit entangling gates via single-photon coupling to Rydberg states \cite{FSQubitPatent,Pagano2022}. Finally, we note that spin-orbit encoding for qubit implementation is also at the heart of various other platforms such as quantum dots \cite{Zwanenburg2013}, color centers \cite{Awschalom2018}, or even large molecular complexes \cite{GaitaArino2019}.

Our experiments start with a single trapped $^{88}$Sr ground state (${^1\rm{S}_0}$) atom, stochastically loaded from a narrow-line magneto-optical trap into an optical tweezer at a wavelength of $\lambda = \unit[539.91]{nm}$ and with a waist of $\unit[564(5)]{nm}$ (Fig.~\ref{Fig1}(b)). For details on trap loading, sideband cooling and parity projection at this wavelength, see our previous work Ref.~\cite{Holzl2023}. After trap loading, we initialize the ${^3\rm{P}_0}$ qubit state via optical pumping by illuminating the atom with three laser beams at about \unit[689]{nm}, \unit[688]{nm}, and \unit[707]{nm} wavelength (\textit{cf.} Fig.~\ref{Fig1}(a)). Typical preparation efficiencies are measured to be about $90 \%$, with state preparation errors attributed to imperfect pumping from the ${^1\rm{S}_0}$ and ${^3\rm{P}_2}$ states. Similarly, we perform state-selective read-out by pumping the population from ${^3\rm{P}_2}$ into the ${^3\rm{P}_1}$ using light at \unit[707]{nm}, from where the atom rapidly decays into the ground state. The branching ratio for the decay from $5\rm{s}6\rm{s} \, ^3\rm{S}_1$ causes partial decay into ${^3\rm{P}_0}$, which limits the fidelity for state-selective read-out in our experiments to below $76 \%$. 

In Fig.~\ref{Fig1}(d), we show coherent Rabi oscillations measured by driving the ${^3\rm{P}_0} \leftrightarrow {^3\rm{P}_2}$ Raman transition with a pair of phase-stable laser beams at about \unit[691.83]{nm} ($\rm{R_1}$) and \unit[665.10]{nm} ($\rm{R_2}$), blue detuned by $\Delta_{\rm{R}}/(2\pi)\approx\unit[9.41]{THz}$ from $^3$S$_1$ and with about \unit[0.5]{mW} (\unit[0.8]{mW}) optical power in $\rm{R_1}$ ($\rm{R_2}$). Both lasers are referenced to a common optical frequency comb, which is in turn stabilized to an ultralow-expansion cavity. The light is focused onto the atom to a Gaussian waist of \unit[6.3]{\um}. Spatial mode matching is granted by sending both beams through the same optical fiber to the experiment. The pair of Raman beams and the optical tweezer beam are counter-propagating and are all linearly polarized along the $y$-direction (\textit{cf.} Fig.~\ref{Fig1}(b)). Along the same direction, we first apply an offset magnetic field $|\vec{B}|$ of $\unit[3]{G}$ ($\phi=0$), which splits the Zeeman substates of the ${^3\rm{P}_2}$ orbital and allows us to selectively address the $m_J=0$ clock level.

\begin{figure}[t]
    \centering
      \includegraphics[width=\columnwidth]{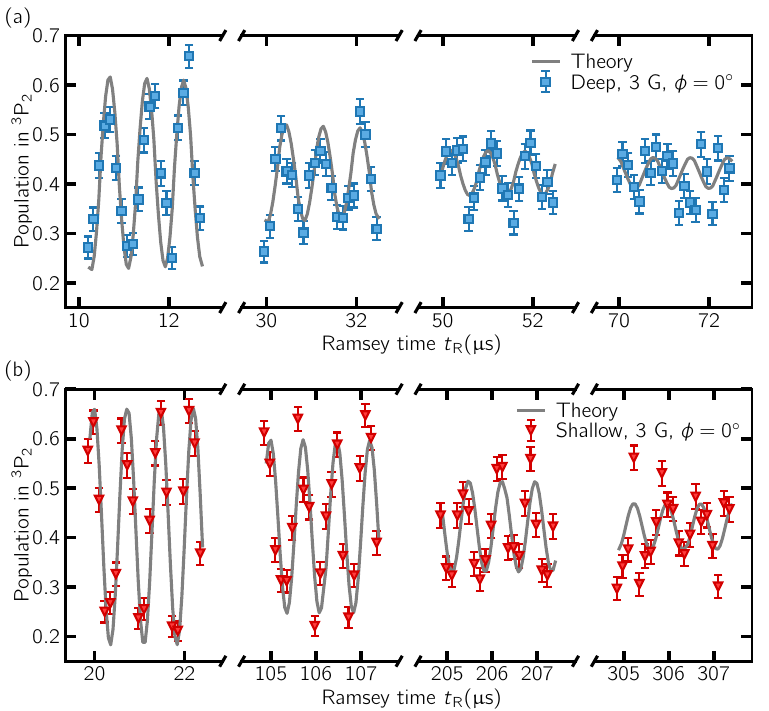}
      \caption{Ramsey interferometry of the fine-structure qubit. Ramsey fringes showing the population in ${^3\rm{P}_2}$ as a function of the time $t_{\rm{R}}$ between the $\pi/2$-pulses as measured in a deep (a) and shallow (b) tweezer with $P=\unit[1.45]{mW}$ and $P=\unit[46]{\upmu W}$, respectively. The magnetic field is set to $|\vec{B}| = \unit[3]{G}$ and $\phi=0^\circ$. Solid gray lines show results from a numerical simulation fit to the data, which takes into account finite temperature.
      }
      \label{Fig2}
\end{figure}

At these conditions, the magic wavelength of the fine-structure qubit, which yields equal trap depth for ${^3\rm{P}_0}$ and ${^3\rm{P}_2}$ (intersection of the solid red and blue line in Fig.~\ref{Fig1}(c)), is about $\unit[4]{nm}$ shorter than our tweezer wavelength \cite{SafronovaPC,Holzl2023}. For the tweezer power set during the Rabi oscillation measurement $P=\unit[1.45]{mW}$, we compute the differential light shift at the trap center $\delta U = U_{{^3}\rm{P}_0} - U_{{^3}\rm{P}_2} = - h \times \unit[0.2]{MHz}$. Even though this yields a substantial non-magicness, we attribute the damping of the Rabi oscillations mainly to shot-to-shot fluctuations of the local Rabi frequency due to position fluctuations of the Raman beams relative to the tweezer. This is supported by a numerical simulation matched to the data (gray line). For this, we time-evolve a laser-driven two-level atom trapped in a 3D harmonic oscillator, taking into account level-dependent trap frequencies computed from the experimental tweezer parameters, finite temperature ($T=\unit[8]{\upmu K}$) and Monte Carlo sampling of the local Rabi frequency. The latter is taken from a Gaussian distribution with a standard deviation of $10 \%$ around a mean value of $\Omega_{\rm{R}} =2\pi \times \unit[84]{kHz}$. This corresponds to large beam position fluctuations (\unit[1.4]{\um}) since we focus the tweezer and Raman beams via independent objectives in the current constrained setup, which can be significantly improved by focusing both via the same objective, enabling fluctuations on the 100 nm level and hundreds of Rabi cycles. 

In a next step, we characterize the transverse coherence time of the qubit via Ramsey spectroscopy, for which we first prepare a coherent superposition of the qubit state by a resonant $\pi/2$-pulse starting from ${^3\rm{P}_0}$. After a variable wait time $t_{\rm{R}}$ we apply a second $\pi/2$-pulse and read out the population in ${^3\rm{P}_2}$. An examplary dataset of the resulting Ramsey oscillations is shown in Fig.~\ref{Fig2}(a) for the same tweezer parameters as set for the Rabi oscillation measurement before. Note that we have slightly adapted the standard Ramsey protocol and force the phase of the second  $\pi/2$-pulse to always start from zero. This is readily implemented experimentally by resetting the phase of two radio frequency signals $f_{\rm{R1}}$ and $f_{\rm{R2}}$, which drive two acousto-optic modulators (double-pass configuration) in the beam paths used for pulsing the individual Raman beams. The phase reset causes the Ramsey phase to evolve at a rate determined by the difference between these frequencies, $2 \lvert f_{\rm{R1}}-f_{\rm{R2}}\rvert \approx \unit[1.3]{MHz}$ , which allows us to observe rapid oscillations much faster than enabled by the $\pi/2$-pulse excitation bandwidth.

\begin{figure}[!t]
  \centering
    \includegraphics[width=\columnwidth]{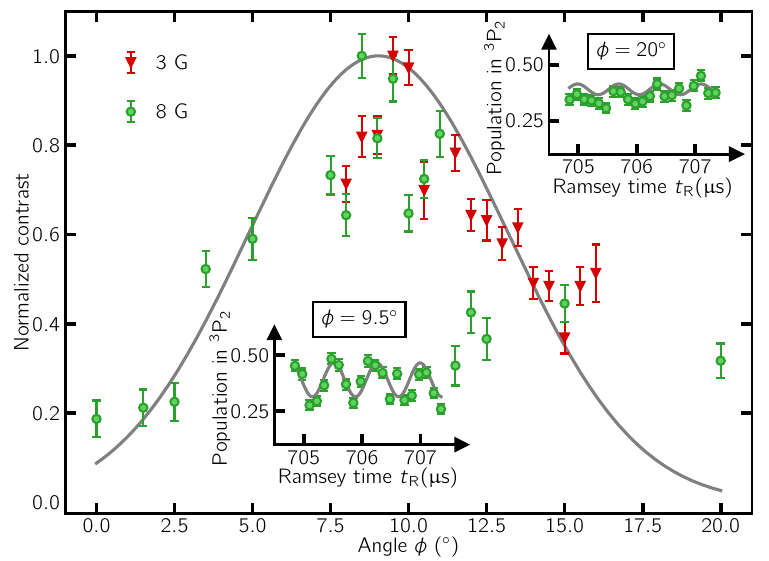}
    \caption{Magic wavelength tuning of the fine-structure qubit. Normalized Ramsey fringe contrast as a function of magnetic field angle $\phi$ for $|\vec{B}| = \unit[3]{G}$ (triangles) and $|\vec{B}| = \unit[8]{G}$ (circles) measured at fixed $t_{\rm{R}} = \unit[600]{\upmu s}$ and $t_{\rm{R}} = \unit[700]{\upmu s}$, respectively. For both datasets $P=\unit[46]{\upmu W}$. The solid line is a Gaussian to guide the eye. Insets show representative datasets for $\phi=9.5 ^\circ$ and $\phi=20 ^\circ$ taken at $|\vec{B}| = \unit[8]{G}$. The contrast is extracted from the amplitude of a sinusoidal fit to the Ramsey oscillations and normalized to the maximum value obtained at the magic angle.}
    
    \label{Fig3}
  \end{figure}

The dominant source of transverse qubit dephasing becomes evident when comparing Fig.~\ref{Fig2}(a) to a measurement recorded at a reduced tweezer power of $P=\unit[46]{\upmu W}$ shown in Fig.~\ref{Fig2}(b), for which we observe coherent oscillations at much longer times. This finding suggests that dephasing is dominated by non-magic qubit trapping in combination with finite temperature \cite{Kuhr2005}. Again our numerical simulations (solid lines in Fig.~\ref{Fig2}(a) and (b)), for which we compute the quantum mechanical time evolution as before, but now incorporating the non-standard Ramsey protocol, strongly support this claim. More specifically, the results in Fig.~\ref{Fig2}(a) and Fig.~\ref{Fig1}(d) assume the same temperature and Rabi frequency fluctuations, while the simulation data in Fig.~\ref{Fig2}(b) use $T=\unit[1.4]{\upmu K}$ as expected for adiabatic ramping to the shallow tweezer setting after trap loading. In summary, the analysis shows that while Rabi oscillations are most sensitive to Raman beam fluctuations, the Ramsey measurements are largely independent of that, and provide an excellent measure for trap-induced qubit dephasing.

\begin{figure}[!t]
  \centering
    \includegraphics[width=\columnwidth]{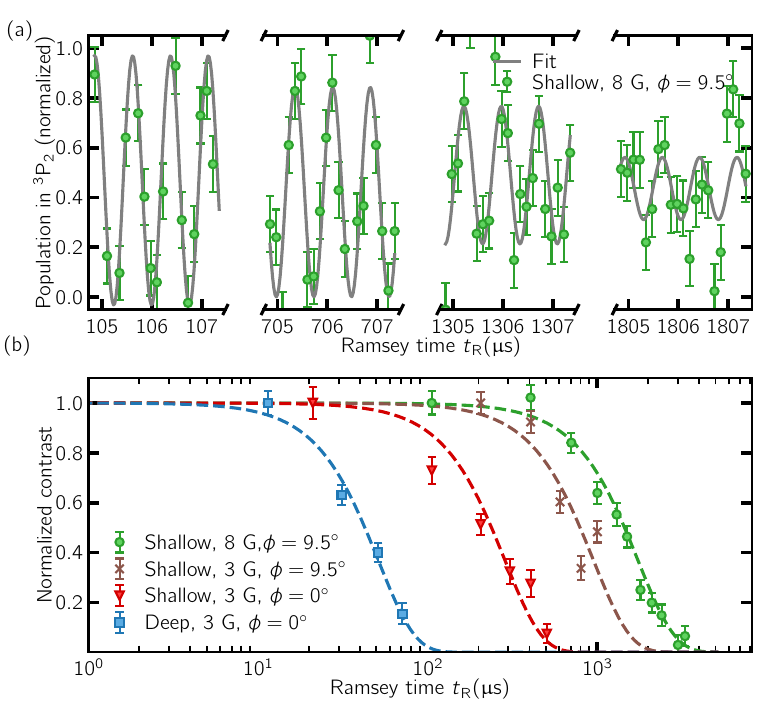}
    \caption{Coherence time of the fine-structure qubit. (a) Ramsey signal normalized to the inital fitted fringe constrast at magic angle $\phi=9.5 ^\circ$ for $|\vec{B}| = \unit[8]{G}$ measured in the shallow tweezer ($P=\unit[46]{\upmu W}$). The gray lines are sinusoidal fits to the data to extract the temporal decay of the contrast. (b) Contrast extracted from fits to data as shown in (a) for $|\vec{B}| = \unit[3]{G},\phi=0^\circ$ in the deep tweezer (squares) and for the shallow tweezer at $|\vec{B}| = \unit[3]{G},\phi=0^\circ$ (triangles), $|\vec{B}| = \unit[3]{G},\phi=9.5^\circ$ (crosses), and $|\vec{B}| = \unit[8]{G},\phi=9.5^\circ$ (circles) with $T_2$ values of 36(2) $\upmu$s, 203(17)$\upmu$s, 676(70) $\upmu$s and 1236(82) $\upmu$s, respectively. Dashed lines show fits based on a Gaussian envelope to extract the qubit coherence time $T_2$. 
    }
    \label{Fig4}
  \end{figure}

To further improve the coherence time, we exploit the tensor polarizability of the ${^3\rm{P}_2}$ state, which allows us to tune the qubit into magic trapping conditions at our tweezer wavelength. Specifically, this is achieved by introducing an angle $\phi = \tan^{-1}(B_x/B_y)$ between the fixed tweezer polarization axis and the external magnetic offset field vector $\vec{B}$, which we rotate in the plane perpendicular to the axial tweezer direction, i.e. in the $x$-$y$-plane (\textit{cf.} Fig.~\ref{Fig1}(b) and (c)). Figure~\ref{Fig3} shows the measured angle dependence of the Ramsey fringe contrast at a fixed value of $t_{\rm{R}}$ for data recorded with $|\vec{B}|=\unit[3]{G}$ and $|\vec{B}|=\unit[8]{G}$. We observe a contrast maximum, indicating optimized magic qubit trapping, at an angle of $\phi \approx \unit[9.5]{^\circ}$. We notice a deviation from the prediction based on polarizability data \cite{SafronovaPC} (\textit{cf.} Fig.~\ref{Fig1}(c)) somewhat larger than estimates for systematic errors on $\phi$ in the experiment.

A Ramsey measurement recorded at this magic trapping angle is shown in Fig.~\ref{Fig4}(a) for $|\vec{B}|=\unit[8]{G}$. Notably, tuning $\phi$ allows us to clearly observe qubit coherence for times exceeding \unit[1.5]{ms}, providing an improvement of more than an order of magnitude compared to the situation presented in Fig~\ref{Fig2}(a). For a more quantitative analysis, we fit the Ramsey data with sinusoidal functions (solid lines), from which we extract the fringe contrast as a function of $t_{\rm{R}}$. The result of that procedure is presented in Fig.~\ref{Fig4}(b) (circles), and is compared to scenarios at $\phi=0^\circ$ for both the shallow and the deep tweezer setting (triangles and squares).

For all datasets, the characteristic transverse coherence time $T_2$ is then quantified by fitting a Gaussian envelope of the form $\exp(-t_{\rm{R}}^2/2T_2^2)$ to the decay of the contrast, reflecting a stochastic dephasing process. Our analysis yields $T_2=\unit[1236(82)]{\upmu s}$ for longest measured coherence time obtained at the magic trapping angle in the shallow tweezer and at \unit[8]{G}. For comparison, at $\phi=0$ we find significantly shorter values of  $T_2=\unit[36(2)]{\upmu s}$ and $T_2=\unit[203(17)]{\upmu s}$ for trapping in the deep and shallow tweezer at 3G, respectively. Finally, a fourth set of data taken at magic angle but at $\unit[3]{G}$ (crosses in Fig.~\ref{Fig4}(b)) yields $T_2=\unit[676(70)]{\upmu s}$, and indicates an improvement in $T_2$ when increasing the magnetic field strength.

\begin{figure}[!t]
  \centering
    \includegraphics[width=\columnwidth]{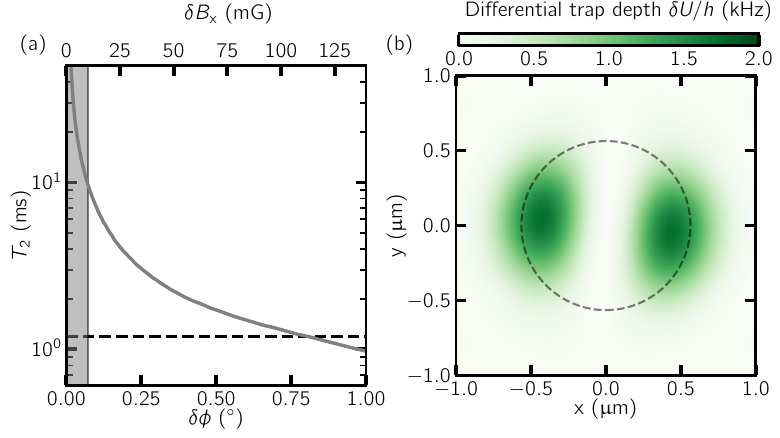}
    \caption{(a) Coherence time $T_2$ predicted from numerical simulations in the presence of shot-to-shot noise in the magnetic field angle $\phi$ of magnitude $\delta \phi$. The upper axis shows the magnitude of transverse magnetic field noise $\delta B_x$ resulting in $\delta \phi$. The shaded region indicates $\delta B_x<\unit[10]{mG}$, and the dashed line depicts the longest  $T_2$ time measured. (b) Differential light shift $\delta U$ through the tweezer focal plane caused by the polarization landscape of the high-NA beam. The dashed circle denotes the $1/e^2$ radius of the tweezer intensity. Results in (a) and (b) are calculated for the experimental parameters of Fig.~\ref{Fig4}(a), i.e. at magic angle and for $P=\unit[46]{\upmu W}$.}
    \label{Fig5}
\end{figure}

Finally, we aim to identify major constraints limiting the observed qubit coherence time in the presence of the tweezer trap and discuss possible improvements for future experiments. To this end, we first note that technical sources causing fluctuations of $\phi$ around the ideal magic condition will affect the achievable $T_2$ time. To analyze the effect of such fluctuations, we include them into our numerical simulations. Specifically, we compute the Ramsey signal for an ensemble of $\phi$-values which are Gaussian-distributed around the magic angle with a standard deviation $\delta \phi$. Averaging these results provides an expected $T_2$ time as a function of $\delta \phi$ (Fig.~\ref{Fig5}(a)). Evidently, fluctuation in $\phi$ may arise from magnetic field noise in the experimental setup. However, the numerical analysis shows that typical noise amplitudes required to explain the best $T_2$ time achieved have to be significantly larger than allowed by the field stability in our experiment, which we have measured independently to be well below \unit[10]{mG}. Furthermore, shot-to-shot fluctuations in the tweezer polarization affecting $\phi$ are expected to be too small to contribute significantly to this effect.

A more fundamental decoherence source stems from the high numerical aperture ($\rm{NA}=0.5$) of the optical tweezer, which causes longitudinal electric field components in the focal spot \cite{Thompson2013}. These components lead to a locally varying reduction of the electric field vector in the $x$-$y$-plane, and consequently a spatial dependence of the tensor light shift contribution in the tweezer. Figure \ref{Fig5}(b) shows the resulting differential light shift $\delta U(x,y)$ in the focal plane ($z=0$) for the experimental parameters of Fig.~\ref{Fig4}(a), i.e. our measurement at magic angle.

The data is obtained by first calculating the complex polarization vector of the tweezer light field $\vec{\epsilon}(x,y)$ throughout the focus \cite{Richards1959,NovotnyBook}. This enters into the ac-Stark Hamiltonian $H$ via \cite{Cooper2018}
$$H/E_0^2 = -\alpha_s - \frac{3\alpha_t}{J(2J-1)} \left( \frac{\{\vec{\epsilon}\cdot \vec{J},\vec{\epsilon}^\ast \cdot \vec{J} \}}{2} - \frac{J(J+1)}{3}\right),$$
where $a_s$ ($a_t$) denotes the scalar (tensor) polarizability, $\vec{J}$ the total angular momentum operator with the associated quantum number $J$, and $E_0(x,y)$ the electric field strength \cite{Cooper2018}. Adding the interaction with the magnetic field $H_B=\mu_B g_J \vec{B}\cdot \vec{J}$ and diagonalizing $H+H_B$ yields light- and Zeeman-shifted energy levels and thus the differential light shift of our qubit for Fig.~\ref{Fig5}(b). 

The influence of the longitudinal field components on $\delta U(x,y)$ yields a quadratic increase with the distance from the trap center along the direction perpendicular to the input polarization vector, before it becomes maximal near the trap waist with a value $\delta U_{\rm{max}}\approx h \times \unit[1.7]{kHz}$. We may estimate the effect on qubit dephasing by averaging $\delta U(x,y)$ over a thermal Gaussian distribution in the focal plane with the temperature obtained previously ($T=\unit[1.4]{\upmu K}$), and find a corresponding timescale of a few milliseconds, in fair agreement with our observed $T_2$ time. We note the possibility to rephase this trap-induced decoherence via spin-echo, for which we expect to reach irreversible coherence times in the range of tens of milliseconds, dictated by heating during the spin-echo protocol \cite{Kuhr2005}.

In conclusion, we have demonstrated a novel fine-structure qubit for neutral atom quantum computing with divalent atoms. We have studied the coherence properties of this qubit in an optical tweezer trap and measured a coherence time of up to \unit[1.2]{ms} when the trap is tuned into magic trapping conditions, exploiting the tensor polarizability of the ${^3\rm{P}_2}$ state. We found that the observed $T_2$ time is limited by the polarization landscape in the tweezer focus. Improved in-trap cooling of the qubit, ideally into the three-dimensional ground state, should significantly increase the coherence time toward the tens of milliseconds range. Additionally, we propose an elegant route to largely eliminate the differential trap depth variations. One may choose a tweezer wavelength for which the magic angle $\phi = \unit[90]{^\circ}$. Considering the focal spot as a superposition of plane waves, it is apparent that the field vectors are all orthogonal to the magnetic field axis, up to the much smaller transverse component. Notably, one finds such conditions in the strontium atom for a wavelength of approximately \unit[755]{nm}, for which we predict a reduction of the maximum differential trap depth as shown in Fig~\ref{Fig5}(b) by two orders of magnitude.

Our target coherence times of tens of milliseconds and longer have to be compared to the exceptionally fast single-qubit rotations, which can be readily achieved via the reported Raman coupling. Specifically, increasing the Raman intensity into the milliwatt regime and using a tighter beam waist enables single-qubit rotations on the \unit[100]{ns} timescale. This provides exciting prospects for achieving $\sim10^5$ gate operations within the qubit coherence time, which is comparable with the so far studied clock  or nuclear spin qubit implementations in divalent atoms.

Finally, high-fidelity readout of the qubit can be implemented in the future via coherent transfer starting from ${^3\rm{P}_0}$, either directly to ${^1\rm{S}_0}$ via a three-photon scheme, or to ${^3\rm{P}_1}$ with two photons, followed by spontaneous decay. Moreover, transfer into a tweezer or lattice system at the clock-magic wavelength at about \unit[813]{nm} \cite{Young2020} would even provide simultaneous magic trapping of the clock- and the fine-structure qubit, enabling coherent state transfer between two qubits encoded in the same atom. In that context the clock transition may serve as an ideal storage with interesting prospects for mid-circuit read-out and feedback in a qudit setting.

In a study performed in parallel to ours, similar results have been achieved with atoms in an optical lattice \cite{Pucher2023}. We thank M. S. Safronova for providing polarizability data, and the QRydDemo consortium and the \href{https://thequantumlaend.de/}{\textsc{Quantum Länd}} team for fruitful discussions. We acknowledge funding from the Federal Ministry of Education and Research (BMBF) under the grants QRydDemo, CiRQus, MUNIQC-Atoms, and the Horizon Europe programme HORIZON-CL4-2021-DIGITAL-EMERGING-01-30 via the project 101070144 (EuRyQa).

%

\end{document}